\documentclass[manuscript]{aastex631}

\usepackage{mathtools} 
\usepackage{bookmark}
\begin{document}

\title{On the Acceleration of the Young Solar Wind from Different Source Regions}

\author[0009-0004-4832-0895]{Yiming Jiao}
\affiliation{State Key Laboratory of Space Weather, National Space 
Science Center, Chinese Academy of Sciences, Beijing, China; liuxying@hotmail.com}
\affiliation{University of Chinese Academy of Sciences, Beijing, China}
\author[0000-0002-3483-5909]{Ying D. Liu}
\affiliation{State Key Laboratory of Space Weather, National Space 
Science Center, Chinese Academy of Sciences, Beijing, China; liuxying@hotmail.com}
\affiliation{University of Chinese Academy of Sciences, Beijing, China}
\author[0009-0005-3941-1514]{Wenshuai Cheng}
\affiliation{State Key Laboratory of Space Weather, National Space 
Science Center, Chinese Academy of Sciences, Beijing, China; liuxying@hotmail.com}
\affiliation{University of Chinese Academy of Sciences, Beijing, China}
\author[0000-0002-8234-6480]{Hao Ran}
\affiliation{Mullard Space Science Laboratory, University College London, Dorking, RH5 6NT, United Kingdom}
\author[0000-0001-5205-1713]{Rui Wang}
\affiliation{State Key Laboratory of Space Weather, National Space 
Science Center, Chinese Academy of Sciences, Beijing, China; liuxying@hotmail.com}
\affiliation{University of Chinese Academy of Sciences, Beijing, China}


\begin{abstract}
    The acceleration of the young solar wind is studied using the first 17 encounters of Parker Solar Probe.
    We identify wind intervals from different source regions: coronal hole (CH) interiors, streamers, 
    and low Mach number boundary layers (LMBLs), i.e. the inner boundaries of coronal holes.
    We present their statistical trends in the acceleration process.
    Most of the observations can be reproduced by a two-fluid hydrodynamic model with realistic corona temperatures.
    In such a model, the solar wind is accelerated by the combined thermal pressures of protons and electrons, 
    but it is mainly the difference in the proton pressure that leads to the difference in the solar wind speed.
    The proton pressure is the highest in the fastest CH wind, with a high initial proton temperature that decreases slowly.
    It is lower in the relatively slow LMBL wind, and the lowest in the slowest streamer wind.
    The proton temperature is quadratically correlated with the wind speed when scaled to the same distance.
    In contrast, the electron temperature shows no significant differences for different wind types or wind speeds, 
    indicating more similar contributions from the electron pressure. 
    The model gives reasonable locations for the sonic critical point, which is on average at 3.6-7.3 solar radii and 
    can also extend to large distances when the proton temperature is extremely low, as in the LMBL wind.
    In addition to the thermal pressure, we raise the possibility that Alfvén waves may contribute to the solar wind acceleration,  
    especially for the fast CH wind.
\end{abstract}



\section{Introduction} \label{sec:intro}

The solar wind acceleration mechanisms have been studied for decades and 
remain an important topic in solar and space physics.
Since the first fluid theory established by \citet{1958ApJ...128..664P}, a number of theoretical works have been devoted to 
explaining how the solar wind is accelerated from the corona 
\citep[e.g.,][]{1960ApJ...132..821P,1967ApJ...148..217W,1968ApJ...151.1155H,1971JGR....76.7479L,1986JGR....91.4111H}.
These early works usually focus on reproducing 1 au observations.
The launch of the Parker Solar Probe \citep[PSP;][]{2016SSRv..204....7F} mission in 2018 
has allowed observations of the solar wind at a much closer distance to the Sun than ever before. 
This provides observational constraints on the radial evolution of the solar wind at its early stage, which allows us to study 
the acceleration of the young solar wind for the first time.

In the studies of the solar wind acceleration, it is typical to divide the wind into fast and 
slow wind according to its speed at 1 au and to consider them separately. 
However, as the solar wind is still accelerating at the close distance of PSP to the Sun, 
dividing it simply by speeds may cause a mixing of streams.
A slow stream near the Sun may be a fast one at 1 au.
Several alternative ways have been developed to define the wind families at PSP distances such as 
the use of the asymptotic speed \citep{2022ApJ...936...53H,2023ApJ...952...26H}, the statistical quantile of speeds 
\citep{2020ApJS..246...62M,2022ApJ...940..130D}, and the correlation of wind speed with proton temperature \citep{2024arXiv240113308P}.
In any case, the differences in the solar wind speed are ultimately rooted in its source region, 
with the fast wind coming from inside the coronal holes \citep[e.g.,][]{1976SoPh...46..303N,1977RvGSP..15..257Z} 
and the slow wind mostly thought to come from closed streamers \citep[e.g.,][]{1997ApJ...484..472S,2009ApJ...691.1936C}. 
Recent works have identified another type of slow wind from PSP measurements that originates from coronal hole inner boundaries, 
termed the low Mach number boundary layer (LMBL) 
\citep{2023ApJ...944..116L,2024ApJ...963...85L,2024ApJ...960...42J,2024ApJ...963...82R,2024ApJ...967...58C}.
Rooted in the rapidly diverging open magnetic fields of the coronal hole boundaries, 
the LMBL wind is characterised by a reduced Alfvén Mach number and leads directly to the first observed sub-Alfvénic solar wind
\citep{2023ApJ...944..116L,2024ApJ...960...42J}, making it a unique component of the pristine solar wind.  
\citet{2024ApJ...960...42J} has shown that the fast coronal hole wind, the streamer wind, 
and the LMBL wind are three main components of the young solar wind. 
In this study, we divide the young solar wind into three families according to its sources. 
The acceleration of these wind families can be then studied by their respective statistical trends.

In the hydrodynamic description, 
the thermal pressure gradient in the young solar wind can cause it to accelerate against the solar gravity.
This process was first established by \citet{1958ApJ...128..664P} using a single-fluid, isothermal hydrodynamic model. 
The model has since been extended in several ways.
One is to represent the protons and electrons as two different species with different temperatures 
\citep[e.g.,][]{1968ApJ...151.1155H,1971JGR....76.7479L,1986JGR....91.4111H,1997SoPh..171..363T,2011ApJ...743..197C}, 
as the weak interaction between these two species of particles can hardly equalise their temperature. 
In such models, the electron pressure acts on the protons through an electric field.
Another is to revise the isothermal assumption that requires infinite energy injection. 
\citet{1960ApJ...132..821P} depicted a model in which the corona obtains enough energy to be isothermal within a distance 
near the Sun and expands adiabatically further away.
Other works tried polytropic models where the plasma get a finite amount of heating 
\citep[e.g.,][]{1964ApJ...139...72P,1964ApJ...139...93P,1999A&A...343..251K,2022PhPl...29l2901S}. 
More recently, \citet{2022ApJ...940..130D} combined the isothermal and polytropic models into an isopoly model.
The model is isothermal out to a certain distance and then followed by a polytropic model,
in which the polytropic indices are determined by in-situ measurements. 
The success of this model in reproducing most of the observations suggests that thermal pressure gradient plays an important role 
in solar wind acceleration.
On this basis, it is necessary to investigate the possible hydrodynamic processes in the three young solar wind families, 
which may reveal key properties of the solar wind source and acceleration regions.


The proton and electron thermal pressure gradients, 
which act to accelerate the solar wind in two-fluid hydrodynamic models, are closely related to the proton and electron temperature profiles.
However, since the protons and electrons are heated from different sources and mechanisms 
\citep[e.g.,][]{1976JGR....81.1649H,2013PhRvL.110i1102K,2017ApJ...839..122K}, 
their temperatures can be very different in the young solar wind and therefore have separate effects on the solar wind acceleration.
This may be reflected in the relationships between the proton and electron temperatures and the velocity.
In fact, while the positive correlation between the proton temperature and solar wind speed is well established 
\citep[e.g.,][]{1987JGR....9211189L,2009SoPh..257..169D,2012JGRA..117.9102E},
the relationship between the electron temperature and speed is less clear. 
Some suggest an anti-correlation \citep{2020ApJS..246...62M,2020ApJS..246...22H,2023A&A...674A..49L,2023ApJ...944...82S},  
but this often depends on the distance \citep{2020ApJS..246...62M,2023A&A...674A..49L,2023ApJ...944...82S}
or wind speed \citep{2023ApJ...944...82S}.
The problem of the temperature-velocity relationship needs to be further discussed for the young solar wind, 
as it may reveal the roles of proton and electron thermal pressures in the wind acceleration.

In this letter, we focus on the acceleration of the young solar wind and examine how different source regions affect the wind properties.
In section \ref{sec:ds},
we show the classification of the solar wind into three sources, including coronal hole interiors, streamers and coronal hole boundaries, 
and present the details of the two-fluid isopoly model we use.
In section \ref{sec:m&a} we show the statistical trends of the three wind types and compare them with the modeling results.
We also discuss how proton and electron temperatures are correlated to the velocities for the three wind families.
The conclusions are summarized in section \ref{sec:c&d}.

\section{Observations and Methodology} \label{sec:ds}

\subsection{Data Classification}

We use PSP in-situ measurements of the first 17 encounters to select solar wind data from different source regions.
Parameters of the solar wind protons are from PSP/SWEAP package \citep{2016SSRv..204..131K} with a combined fields of view of 
the Solar Probe Cup \citep[SPC;][]{2020ApJS..246...43C} and the Solar Probe Analyzer-Ions \citep[SPAN-I;][]{2022ApJ...938..138L}.
The alpha particle data are also available starting from encounter 4.
The magnetic field data are obtained from the measurements of the PSP/FIELDS fluxgate magnetometer instrument \citep{2016SSRv..204...49B}. 
The electron density and core temperature are derived from quasi-thermal noise \citep[QTN;][]{2020ApJS..246...44M} measurements.
As the velocity distribution function can be partially out of the field view for some time, 
the QTN electron density is used as a proxy for the plasma density throughout the study, 
and the proton temperature should be considered as a lower limit.
All parameters are interpolated to 1 minute averages. 

\begin{figure}
\includegraphics[width=\textwidth]{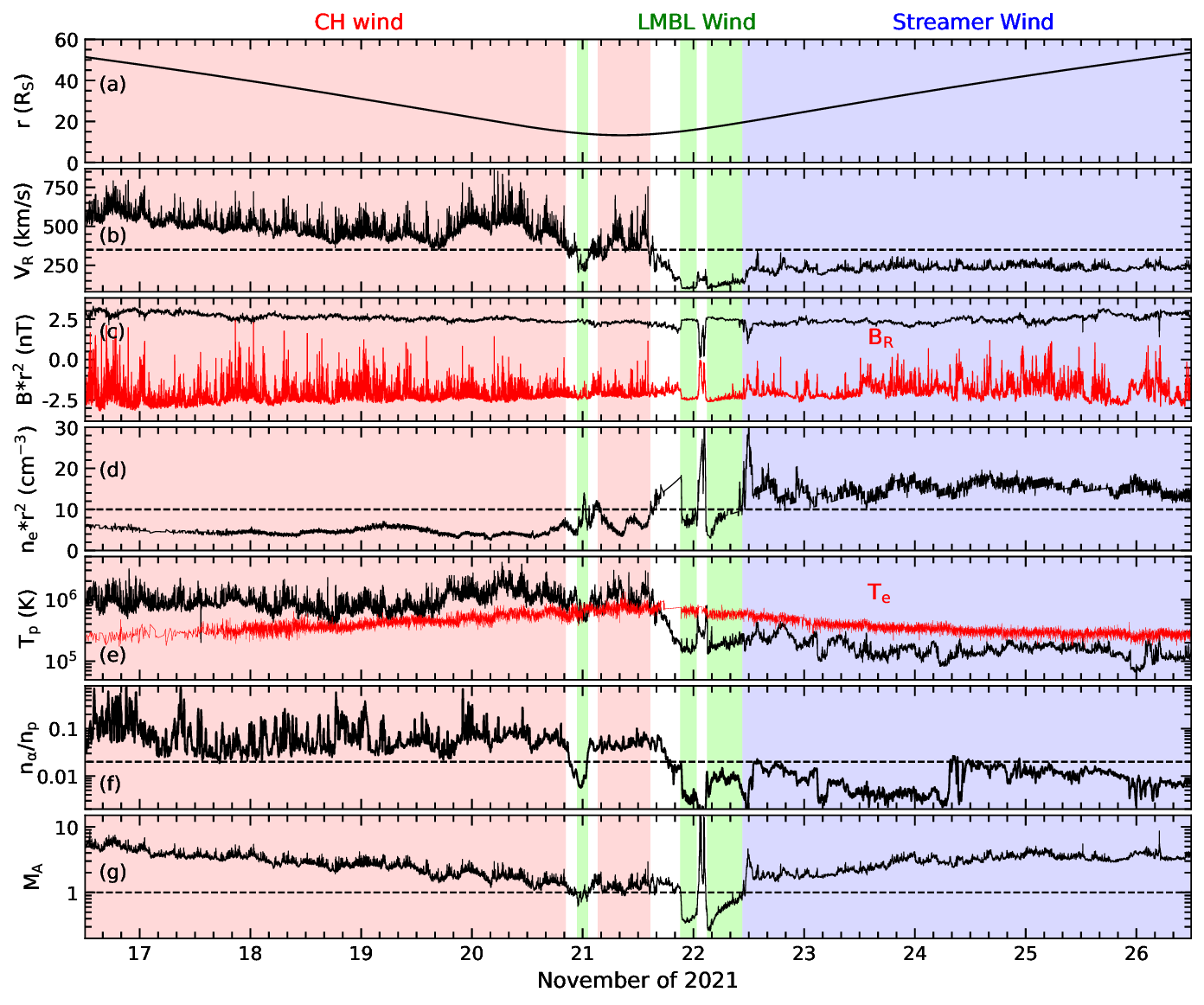}
\caption{PSP measurements at encounter 10 as an example of data classification. 
The shaded areas indicate coronal hole wind (red), streamer wind (blue) and LMBL wind (green). 
(a) PSP heliocentric distance. 
(b) Proton radial velocity. 
(c) Magnetic field strength and radial component (red) normalized to 1 AU values. 
(d) QTN electron density normalized to 1 AU values. 
(e) Proton temperature and electron core temperature (red).
(f) Alpha-to-proton density ratio. The horizontal dashed line marks the value of 0.02.
(g) Radial Alfvén Mach number. }
\label{fig:1}
\end{figure}

As an example of the data selection, Figure \ref{fig:1} shows the measurements from encounter 10.
The basic criteria for coronal hole (CH) wind and streamer wind are that CH wind typically has a higher velocity above 350 km/s 
and a lower normalized density below 10 cm$^{-3}$, 
whereas streamer wind has a lower velocity below 300 km/s and a higher normalized density above 10 cm$^{-3}$. 
The threshold values are chosen based on the general properties of the solar wind at PSP distances 
\citep[e.g.,][]{2023ApJ...944..116L,2024ApJ...960...42J}.
These two types of wind are identified in Figure \ref{fig:1} with corresponding features of the velocity $V_R$ (Figure \ref{fig:1} (b)) 
and the electron density $n_e$ (Figure \ref{fig:1} (d)).
The electron density is normalized to 1au by a $1/r^2$ factor, 
with $r$ being the heliocentric distance (Figure \ref{fig:1} (a)).  
Moreover, previous works suggest that a difference in the alpha-to-proton density ratio, 
determined by the wind sources, induces differences between CH and streamers.
In general, $n_\alpha/n_p$ is higher in the fast CH wind than in the slow streamer wind 
\citep{2001GeoRL..28.2767A,2007ApJ...660..901K,2012ApJ...745..162K,2018MNRAS.478.1884F}, 
consistent with the intervals we have identified (Figure \ref{fig:1} (f)).
We use $n_\alpha/n_p$ above or below 0.02 as a secondary criterion to identify the CH wind and the streamer wind.

For the identification of the solar wind coming from the coronal hole boundaries, or the LMBL wind, 
we adopt the criteria originally proposed by \citet{2023ApJ...944..116L} 
that include a decreased radial Alfvén Mach number, a reduced density, and a relatively low velocity. 
We choose intervals where in general $V_R$ is below 350 km/s, $n_e$ is below 10 cm $^{-3}$, 
and $M_A$ is lower than its neighboring wind as the LMBL.
The radial Alfvén Mach number $M_A$ is given by $V_R\sqrt{\mu n_e m_p}/B$, 
where $\mu$ is the vacuum magnetic permeability,  $m_p$ is the proton mass, and $B$ is the magnetic field strength.
The identified LMBL intervals are also shown in Figure \ref{fig:1}.
In these inervals, a decreased $M_A$ (Figure \ref{fig:1} (g)) is naturally resulted from a reduced $n_e$ and a relatively low $V_R$, 
as $B$ tends to remain roughly constant (Figure \ref{fig:1} (c)).
Moreover, these LMBL intervals are located in a transition region from the CH wind to the streamer wind, 
which is consistent with their nature as boundary layers.

Based on the above criteria, we identify the solar wind intervals of the three sources from each encounter of PSP measurements. 
Another effective way to determine the source of a wind interval is to associate it with magnetic connectivity or EUV imaging. 
However, due to the large amount of data involved in our study, we instead use the given selection criteria for simplicity.

Table \ref{tab:1} lists the time durations of the three types of wind we identified from each encounter so far.
Encounter 8 is not included due to a lack of QTN measurements of the electron core temperature, 
and encounter 11 is not included due to a lack of QTN data at all.
We are cautious with the CH and streamer intervals for the first three encounters, 
as the alpha particle density is not available and cannot be used as a selection criterion.
The total durations of the three types of wind are summarized in the last row of Table \ref{tab:1}.
These durations are hundreds of hours for each wind type and cover the young solar wind from 14 to 54 solar radii, 
providing a statistical basis for studying the radial evolution of the young solar wind.

\subsection{Two-fluid Isopoly Model}

We investigate the possible hydrodynamic processes using a spherically symmetric two-fluid isopoly model, 
similar to those of \citet{2022ApJ...940..130D} and \citet{2022PhPl...29l2901S}.
The equations of conservation of mass and momentum are 
\begin{equation}
    n_e\,V_R\,r^2=constant,\label{eq:1}
\end{equation}
\begin{equation}
     V_R\,\frac{dV_R}{dr}=-\frac{1}{n_em_p}\sum_{s=p,e}\frac{dP_s}{dr}-\frac{G\,M}{r^2},\label{eq:2}
\end{equation}
where $P_s$ is the pressure of the particle species $s$, which is either protons $p$ or electrons $e$, 
$G$ is the gravitational constant, and $M$ is the solar mass.
The pressure $P_s$ is related to the temperature $T_s$ by the ideal gas law $P_s=n_ek_BT_s$, where $k_B$ is the Boltzmann constant.
Integrating Equation \ref{eq:2} from a given point $r_0$ to any point $r$ we have 
\begin{equation}
    \frac{1}{2}[V_R(r)^2-V_R(r_0)^2]=\sum_{s=p,e}\Delta W_{s}+G\,M(\frac{1}{r}-\frac{1}{r_0}),\label{eq:3}
\end{equation}
where $\Delta W_{s}$ is the work per unit mass done by the pressure gradient force of species $s$ from $r_0$ to $r$.
For each particle species $s$, the isopoly model contains an isothermal regime in the solar vincinity out to the distance $r_{\rm{iso},s}$ 
and a polytropic regime further away.
In the isothermal regime $r<r_{\rm{iso},s}$, the temperature $T_s(r)=T_s(r_0)$ is constant and the pressure gradient can be written as 
\begin{equation}
    \frac{dP_s}{dr}=k_B\,T_s(r_0)\,\frac{dn_e}{dr}.\label{eq:4}
\end{equation}
In the polytropic regime $r>r_{\rm{iso},s}$, the polytropic relation gives $T_s\,n_e^{-(\gamma_s-1)}=constant$, 
where $\gamma_s>1$ is the polytropic 
index, and $\gamma_s=1$ corresponds to the isothermal case.
In the form of the power-law evolution $T_s\propto r^{\alpha_s}$ and $n_e\propto r^{\beta}$, 
the polytropic index is calculated as $\gamma_s=1+\alpha_s/\beta$.
The pressure gradient is then obtained using the polytropic relation as
\begin{equation}
    \frac{dP_s}{dr}=\frac{\gamma_s}{\gamma_s-1}n_e\,k_B\,\frac{dT_s}{dr},\label{eq:5}
\end{equation}
Plugging Equation \ref{eq:4} and Equation \ref{eq:5} into the pressure gradient term of Equation \ref{eq:2} and integrating it, 
in combination with the polytropic relation and Equation \ref{eq:1}, we get
\begin{equation}
    \Delta W_{s}=
    \begin{dcases} 
        -c_s(r_0)^2\,ln[\frac{V_R(r_0)\,r_0^2}{V_R(r)\,r^2}],& \gamma_s=1\\[3mm]
        -\frac{c_s(r_0)^2}{(\gamma_s-1)}\,([\frac{V_R(r_0)\,r_0^2}{V_R(r)\,r^2}]^{\gamma_s-1}-1),& \gamma_s>1,
    \end{dcases}\label{eq:6}
\end{equation}
where $c_s(r_0)^2=\gamma_sk_BT_s(r_0)/m_p$ is the square of the equivalent sound speed of species $s$ at $r_0$. 
This means that in any regime of the isopoly model, as long as $T_s(r_0)$ and $V_R(r_0)$ at a given point $r_0$ are known,
the velocity $V_R(r)$ at any other point $r$ can be solved from Equation \ref{eq:3} and Equation \ref{eq:6}. 
Then $n_e(r)$ is given by Equation \ref{eq:1} and $T_s(r)$ is given by the polytropic relation.

Since the protons and electrons transit from isothermal to polytropic at different distances $r_{\rm{iso},p}$ and $r_{\rm{iso},e}$, 
the model actually contains three layers.
The first layer is the isothermal regime for both species that $\gamma_p=\gamma_e=1$.
In this case, Equation \ref{eq:3} and Equation \ref{eq:6} can be simplified to the isothermal solar wind model in 
\citet{1958ApJ...128..664P} that passes through a singular point
\begin{equation}
    r_c=\frac{G\,M}{2\,V_c^2} \quad \mbox{and} \quad V_c^2=\frac{k_B\,(T_{p0}+T_{e0})}{m_p}, \label{eq:7}
\end{equation}
which serves as the known point $r_0$ when the initial temperatures $T_{p0}$ and $T_{e0}$ are given.
The point $r_c$ is the sonic critical point where the solar wind transitions from subsonic to supersonic, and $V_c$ is the wind speed at $r_c$.
We set initial temperature $T_{s0}$ and isothermal distance $r_{\rm{iso,s}}>r_c$ for each species $s$ as the input parameters.
We assume that the sonic critical point occurs in the first layer \citep[see][]{2022ApJ...940..130D}. 
Then the first layer of the model is solved with $r_c$ as the given point.
The second layer is isothermal for one species and polytropic for the other.
For example, if $r_{\rm{iso},p}<r_{\rm{iso},e}$, 
the second layer is characterized as $r_{\rm{iso,p}}<r<r_{\rm{iso,e}}$, $\gamma_e=1$, and $\gamma_p>1$.
Since the model is continuous, this layer can be solved by using $r_{\rm{iso,p}}$, the end point of the previously solved layer, 
as the known point $r_0$.
The last layer in our example has $r>r_{\rm{iso},e}$, and it is polytropic for both species. 
It can be solved by using the end point of the second layer $r_{\rm{iso},e}$ as $r_0$.
Thus we have a complete set of profiles of $V_R$, $n_e$, $T_p$ and $T_e$ relative to $r$ given by the model.

\section{Comparison between Observations and Model} \label{sec:m&a}

\begin{figure}
    \includegraphics[width=\textwidth]{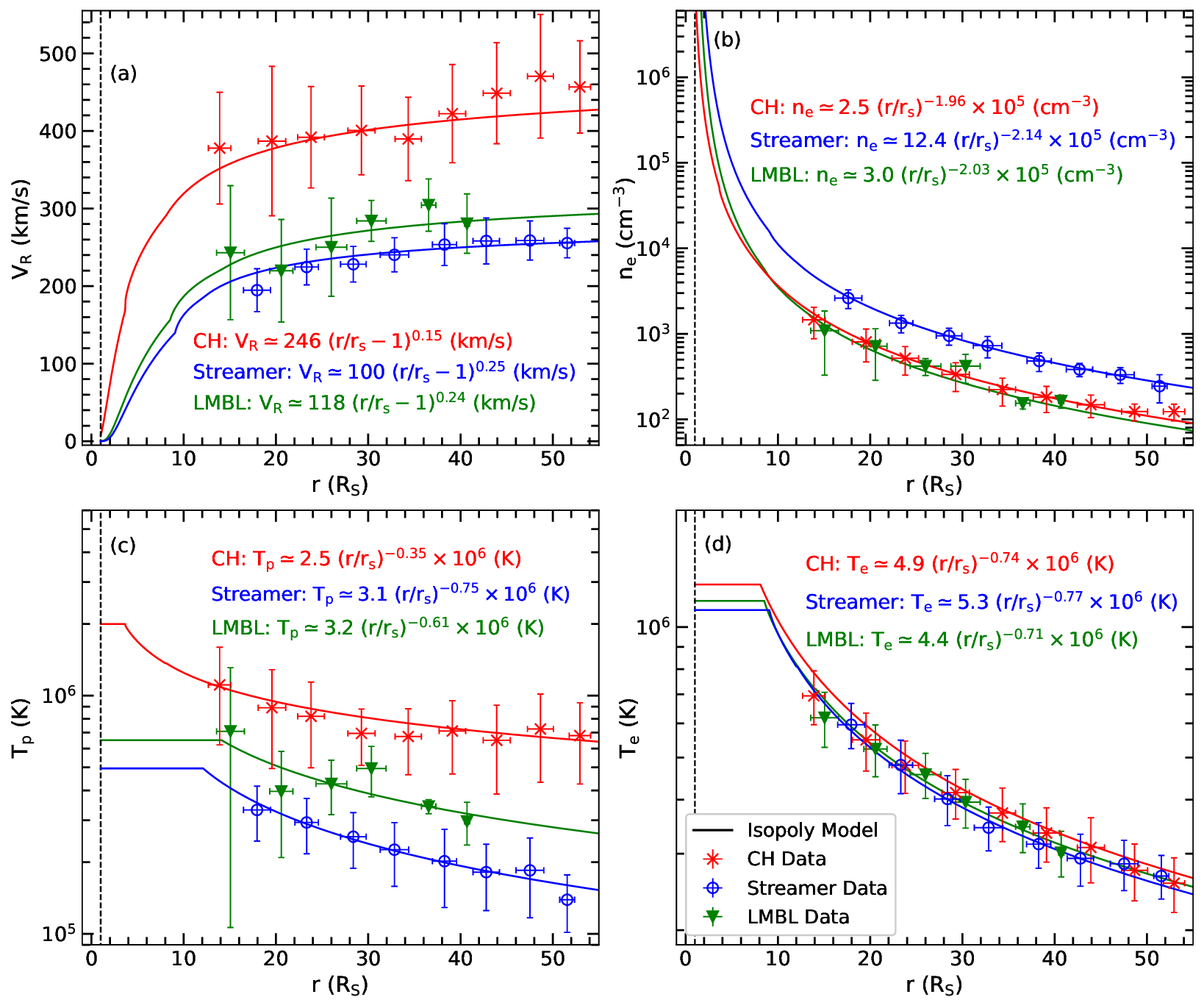}
    \caption{Radial evolution of the solar wind parameters.
    (a) Radial velocity.
    (b) Electron density.
    (c) Proton temperature.
    (d) Electron temperature.
    The PSP data are binned every 5 $R_S$, 
    and the data points and error bars are the mean values and standard deviations within the bins.
    The fitted power-law functions are shown as text.
    The solid lines indicate profiles given by our isopoly model.}
    \label{fig:2}
    \end{figure}

Figure \ref{fig:2} presents the variations of velocity $V_R$, density $n_e$, 
proton temperature $T_p$ and electron temperature $T_e$ as a function of heliocentric distance $r$.
We bin the data every five solar radii and calculate the mean values and standard deviations of $r$ and of each parameter. 
These values are shown as data points and error bars.
We notice an inconsistency between the $T_p$ values measured by SPC and by SPAN-I that $T_p$ from SPC is much lower.
This could be explained by the fact that the SPC measurements are dominated by the core of the proton distribution function, 
while the SPAN-I measurements also include the nonthermal component \citep{2022ApJ...940..130D}.
As SPAN-I measurements alone can cover most of the radial range during the encounters, 
only $T_p$ from SPAN-I is used in the following study.

We first fit the data points with a power-law function to describe their trends empirically.
For  $n_e$, $T_p$ and $T_e$, such a form is efficient in describing the expansion process.
For $V_R$, it is used to represent the acceleration of the solar wind, 
which is rapid near the Sun and then slower further away. 
We set the acceleration to start from the solar surface.
The fitted formulas are shown in Figure \ref{fig:2}, but their associated curves are not.
The fitted expressions for $V_R$ show the acceleration for the three wind types, 
and the contrast between the fast wind coming from CHs, the slow wind coming from streamers and the LMBL wind, 
with the streamer wind being the slowest. 
The CH wind here is below the typical speed of the fast wind at 1 au, 
because the wind is still accelerating near the Sun. 
Also, the wind coming from the central parts of large coronal holes is rarely measured by PSP so far.
The lower speed limit set in the CH wind selection criteria may also have a small effect on this result.
The fitting of $n_e$ shows trends that do not deviate far from the spherical expansion of $n_e\propto r^{-2}$, 
and its values are lower for the CH and LMBL wind than for the streamer wind.
As for $T_p$, it decreases with $r$ at different rates for the three wind types: 
most rapidly for the streamer wind, slower for the LMBL wind, and slowest for the CH wind. 
The initial values in the fitting are similar for the three wind types. 
However, as the trends may not extend to the Sun, they cannot be regarded as true coronal values. 
The expressions for $T_e$ are consistent with the previously estimated $T_e \propto r^{-0.74}$ trend \citep{2020ApJS..246...44M}, 
and show no significant differences between different wind types.
Both $T_p$ and $T_e$ decrease slower than an adiabatic expansion $T \propto r^{-4/3}$ for all wind types, 
suggesting universal heating in the near-Sun environment. 

The profiles given by our isopoly model that fit the measured data point are also shown in Figure \ref{fig:2}.
For each wind type, given the power-law indices for $T_p$, $T_e$ and $n_e$ from in situ fitting, 
the polytropic indices $\gamma_p$ and $\gamma_e$ are obtained. 
The initial temperatures $T_{p0}$ and $T_{e0}$ and the isothermal distances $r_{\rm{iso,p}}$ and $r_{\rm{iso,e}}$ 
are manually adjusted to fit the data.
Since $r_{\rm{iso,s}}$ depends on $T_{s0}$ to ensure a continuous transition from isothermal to polytropic, 
the only free parameters are $T_{p0}$ and $T_{e0}$. 
Most of the observations can be reproduced with realistic parameters, 
except for the CH wind beyond 40 solar radii (Figure \ref{fig:2} (a)), where the model predicts a slightly lower $V_R$ than the measurements.
This discrepancy suggests the possibility of an acceleration mechanism other than the thermal pressure in the CH wind.
The sonic critical point $r_c$ and the critical speed $V_c$ are given by the initial temperatures using Euqation \ref{eq:7}.
The Alfvén critical point $r_A$, which is the point where the Alfvén speed $V_A=B/\sqrt{\mu\,n_e m_p}$ equals the solar wind speed $V_R$, 
is determined by finding the intersection of the two curves of $V_R$ and $V_A$, 
where $V_R$ and $n_e$ are obtained from the model and $B$ is from a power-law fitting of the measurements.

Table \ref{tab:2} summarises the model parameters for the three wind types that relate to the profiles in Figure \ref{fig:2}.
The proton polytropic index $\gamma_p$ is the smallest for the CH wind at 1.18 and the largest for the streamer wind at 1.35, 
with the LMBL wind in between at 1.30.
The much lower $\gamma_p$ for the CH wind suggests that 
the CH wind is most approximate to isothermal and obtains most heating and acceleration.
The initial proton temperatures $T_{p0}$ varies from 0.5 Mk to 2.0 Mk,  
with the streamer wind having the lowest value and the CH wind the highest.
Both small $\gamma_p$ and high $T_{p0}$ produce a higher $T_p$ profile and greater acceleration in the CH wind, 
making it faster than the streamer and LMBL wind (see Figure \ref{fig:2}). 
On the contrary, for electrons both the polytropic indices $\gamma_e$, which are about 1.36, and the initial temperatures $T_{e0}$, 
which are a little more than 1 Mk, are similar for the three wind types.
This yields similar $T_{e}$ profiles for the three wind types as is the case of the fitted power-law expressions (see Figure \ref{fig:2}).
These results suggest that it is mainly the differences in the proton temperatures that cause the different speeds for different wind types. 
The sonic critical point positions $r_c$ are slightly higher than previous estimations 
\citep[e.g.,][]{1995GeoRL..22.1465H,1997ApJ...484..472S,2003ApJ...598.1361V}.
Specifically, according to Equation \ref{eq:7}, $r_c$ is closer to the Sun at higher temperatures, 
so in general the CH wind has the smallest $r_c$ and the streamer wind the largest.
This pattern can sometimes be broken, 
as the LMBL wind can have an extremely low proton temperature and thus an extremely extended $r_c$ 
\citep{2024ApJ...963...85L,2024ApJ...967...58C}. 
The Alfvén critical points $r_A$ are around 12 solar radii for the CH wind and the streamer wind, 
and rise to over 20 solar radii for the LMBL wind, consistent with estimations from PSP in situ measurements 
\citep{2021ApJ...908L..41L,2023ApJ...944..116L,2024ApJ...960...42J}.

\begin{figure}
    \includegraphics[width=\textwidth]{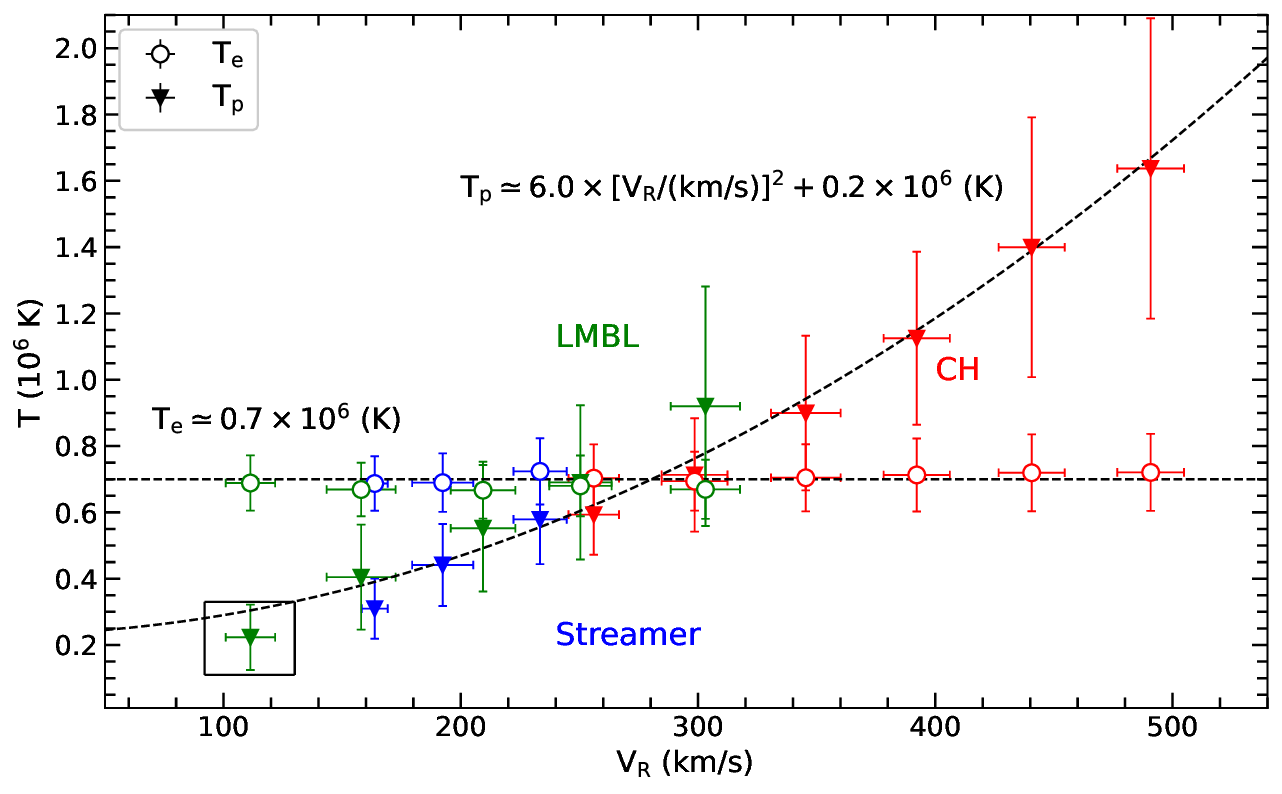}
    \caption{Relationships of proton and electron temperatures with the radial velocity.
    The $V_R$, $T_p$ and $T_e$ data are scaled to the same distance (14 $R_S$) and binned every 50 km/s.
    The dashed lines indicate the fitted functions for protons and electrons, respectively.
    The box marks the LMBL wind with extremely low velocity and an extremely low proton temperature.}
    \label{fig:3}
    \end{figure}

We further investigate the roles of proton and electron temperatures in solar wind acceleration 
by looking at their relationships with the velocities.
To eliminate the effects of radial variations in velocity and temperatures, 
we scale $V_R$, $T_p$ and $T_e$ to 14 $R_S$, the outermost isothermal distance of the isopoly model, 
with the previously fitted power-law functions.
We bin the data every 50 km/s of $V_R$ and calculate the mean values and standard deviations of $V_R$, $T_p$ and $T_e$ within each bin.
The results are shown in Figure \ref{fig:3}.
The scaled $T_p$ is positively correlated with $V_R$, 
as expected from the intuition that a higher temperature produces more acceleration.
This positive correlation can be described by a quadratic function for all the three wind types as shown in Figure \ref{fig:3}, 
which implies a linear relationship between the thermal energy and kinetic energy.
On the other hand, the scaled $T_e$ is around a constant value of 0.7 MK independent of $V_R$.
It appears that after removing the effects of the radial variations in $V_R$ and $T_e$, 
the electon temperature is approximately the same for each wind speed and wind type.
The difference in the relationships of $T_p$ and $T_e$ to $V_R$ is also shown in the in situ measurements in Figure \ref{fig:1} (e), 
where $T_p$ varies significantly with $V_R$ but $T_e$ does not.
This again suggests that, as the electron thermal pressure contributes similarly to the acceleration of all solar winds, 
it is mainly the proton pressure that determines the differences in the wind speed.
In general, the CH wind has the highest $T_p$ and $V_R$ values, the streamer wind has the lowest, 
and $r_c$ is in the same order from the closest to farthest.
However, the variation in the $T_p$ and $V_R$ of the LMBL wind is so large that they can reach extremely low values as 
marked in Figure \ref{fig:3}.
This can lead to an extremely extended $r_c$ in some LMBL winds.
This corresponds to the first observed near-subsonic LMBL wind featuring an extremely low velocity and an extremely low proton temperature 
\citep{2024ApJ...963...85L,2024ApJ...967...58C}.

Accelerating the solar wind requires energy from a variety of sources.
\citet{2023ApJ...952...26H} quantified the energy budget and found that 
the electric field and Alfvén waves provide enough energy to accelerate the slow and fast solar winds.
We repeat their procedures for the three types of wind and get consistent results on the energy budget (not shown here).
In the acceleration of the streamer wind, the electric field, 
which is equivalent to the electron pressure gradient \citep[see][]{2010AIPC.1216....3P}, provides most of the energy required.
This is also reflected in Table \ref{tab:2} that $T_{e0}$ is much higher than $T_{p0}$ for the streamer wind and LMBL wind. 
Figure \ref{fig:3} also suggests that the slower the wind, the higher $T_e$ than $T_p$. 
In the CH wind, in addition to the electric field, 
the Alfvén waves can provide the extra kinetic energy as needed for further acceleration.
Alfvén waves can both heat the protons and directly accelerate the solar wind \citep[e.g.,][]{1973ApJ...181..547H}.
This may explain the greater heating in the CH wind protons, as indicated by the higher $T_{p0}$ and lower $\gamma_p$ 
as well as the additional acceleration in the CH wind that the model cannot produce.
Alfvén wave energy is also present in the LMBL wind acceleration, although to a lesser extent than in the CH wind.

\section{Conclusions} \label{sec:c&d}

We have studied the acceleration of the young solar wind from different source regions using the first 17 encounters of PSP measurements.
We identify the CH wind, the streamer wind, and the LMBL wind as the three main components of the young solar wind, 
and analyze their statistical trends in the velocity, density, and proton and electron temperatures with distance.
The main conclusions are as follows:
\begin{enumerate}
\item[1.] 
The acceleration process of the three types of solar wind can mostly be described by a two-fluid isothermal model.
The velocity and density profiles, which are consistent with the previously proposed fast and tenuous CH wind, 
slow and dense streamer wind, and relatively slow but tenuous LMBL wind \citep{2023ApJ...944..116L,2024ApJ...960...42J}, 
suggest a near-spherical expansion with a small amount of acceleration in the range of the PSP encounter measurements.
The proton and electron temperature profiles show a polytropic decrease with distance.
When appropriate isothermal layers are added before the polytropic region \citep{2022ApJ...940..130D,2022PhPl...29l2901S}, 
the proton and electron thermal pressures provided by such temperature profiles are sufficient to reproduce most of the observed velocities 
by a hydrodynamic mechanism.
The streamer and LMBL wind can be fully generated with realistic initial corona temperatures, 
and the CH wind can be largely explained except for a slightly additional accelerated part beyond 40 solar radii.
This indicates the important role of thermal pressure in accelerating the solar wind.
\item[2.]  
The proton and electron temperatures and their corresponding thermal pressures have different effects on the acceleration of the young solar wind.
The difference in the proton thermal pressure is primarily responsible for the difference in the wind speed.
For the fast CH wind, both a high initial temperature and a subsequent small polytropic index indicate more heating of the protons 
than in other wind types, and thus more proton pressure contributing to the acceleration.  
In comparison, the proton pressure that can accelerate the relatively slow LMBL wind is much less, 
as indicated by a lower initial temperature and a larger polytropic index, 
and the proton pressure that can accelerate the slowest streamer wind is the least.
This is also supported by the positive correlation between the scaled proton temperature and wind speed, 
which can be described by a quadratic function that holds for all wind types. 
In contrast, the electron thermal pressure in our results shows 
less difference in its contribution to the acceleration for different wind types.
This is observed in the similar electron initial temperatures and similar electron polytropic indices for all wind types.
In addition, the electron temperature is roughly the same regardless of the wind speed or wind type when scaled to the same distance.
If there is no scaling, the radially increasing velocity and the decreasing electron temperature may exhibit an anti-correlation, 
which should diminish when the measurements are made over a short distance or at a location far enough away that 
the wind speed is no longer increasing. 
The different roles of the proton and electron temperatures in the solar wind acceleration may be due to separate heating processes 
for the two species in the corona.
This may require more detailed investigation in the future.
\item[3.] 
The model gives reasonable positions of the sonic critical point and the Alfvén critical point.
The position of the sonic critical point is determined by the proton and electron temperatures together.
The higher the temperatures, the closer the sonic critical point to the Sun.
Since the electron temperatures are similar, 
the critical point positions of the three wind types are generally in the descending order of the proton temperatures, 
i.e. first the CH wind at 3.6 $R_S$, then the LMBL wind at 6.2 $R_S$, and finally the streamer wind at 7.3 $R_S$.
However, since the proton temperature of the LMBL wind can also reach very low values, 
its critical point can extend to large distances and even be detected by PSP \citep{2024ApJ...963...85L,2024ApJ...967...58C}. 
We also obtain the position of the Alfvén critical point consistent with previous estimations, 
which is about 12 solar radii for the CH wind and the streamer wind, but over 20 solar radii for the LMBL wind 
\citep{2023ApJ...944..116L,2024ApJ...960...42J}.
\item[4.]
The acceleration of different types of wind is driven by energy from different sources.
The slow solar wind, especially the streamer wind, where the proton pressure contributes little to the acceleration, 
is mainly driven by the electron pressure.
In this case, the electrons act on the solar wind through an electric field, 
which converts the electrical potential energy of the solar wind into kinetic energy.
On the other hand, in the CH wind, where the proton pressure contributes more to the acceleration, 
another source of energy can heat the protons and thus accelerate the solar wind.
The Alfvén waves abundant in the CH wind are a promising candidate, 
as the energy they contain is sufficient for the acceleration \citep{2023ApJ...952...26H}.
Moreover, the Alfvén waves can directly accelerate the solar wind in addition to heating the protons, 
which may be relevant for the additional acceleration in the CH wind beyond 40 solar radii.
The fewer Alfvén waves in the LMBL wind may also act to heat the protons and accelerate them to higher speeds than in the streamer wind.
The role of the Alfvén waves is a potentially helpful point to consider in further refinement of the model.
\end{enumerate}

The research was supported by the Strategic Priority Research Program of the Chinese Academy of Sciences (No.XDB 0560000), 
by NSFC under grant 12073032,
by the National Key R\&D Program of China No. 2022YFF0503800 and No. 2021YFA0718600, 
and by the Specialized Research Fund for State Key Laboratories of China.
H.R. is supported by a studentship from the UK Science and Technology Facilities Council (STFC). 
We acknowledge the NASA Parker Solar Probe mission and the SWEAP and FIELDS teams for the use of data. 


\bibliography{sample631}{}
\bibliographystyle{aasjournal}



\clearpage

\begin{deluxetable}{cccc}
\tabletypesize{\small}
\tablewidth{100pt} 
\tablecaption{Durations of Solar Wind Intervals from Different Sources\label{tab:1}}
\tablehead{
\colhead{Enc.} & \colhead{CH wind}&\colhead{Streamer wind} & 
\colhead{LMBL wind}\\
\colhead{} & \colhead{(hr)}&\colhead{(hr)} & \colhead{(hr)}
} 
\startdata 
1&27.7&43.2&0\\
2&23.7&41.0&54.3\\
3&34.3&19.8&0\\
4&57.0&116.7&10.5\\
5&6.5&47.8&11.5\\
6&46.7&6.2&19.2\\
7&45.0&131.5&4.5\\
9&33.7&23.3&12.2\\
10&115.0&97.0&12.5\\
12&104.7&51.2&43.3\\
13&112.4&10.7&27.2\\
14&130.5&29.0&21.7\\
15&71.7&0&20.2\\
16&38.8&5.5&31.0\\
17&25.7&5.5&8.3\\
\hline
Total&817.2&635.8&280.0\\
\enddata
\tablecomments{Durations of the CH wind, streamer wind and LMBL wind identified from the first 17 encounters of PSP measurements. 
Encounters 8 and 11 are not included due to incomplete or missing QTN measurements.}
\end{deluxetable}

\begin{deluxetable}{cccc}
\tabletypesize{\small}
\tablewidth{0pt} 
\tablecaption{Parameters Related to the Isopoly Model\label{tab:2}}
\tablehead{
\colhead{} & \colhead{CH wind}&\colhead{Streamer wind} & \colhead{LMBL wind}
} 
\startdata 
$\alpha_p$&-0.35&-0.75&-0.61\\
$\alpha_e$&-0.74&-0.77&-0.71\\
$\beta$&-1.96&-2.14&-2.03\\
$\gamma_p$&1.18&1.35&1.30\\
$\gamma_e$&1.37&1.36&1.35\\
\hline
$T_{p0}$ (MK)&2.00&0.50&0.65\\
$r_{\rm{iso,p}}$ ($R_S$)&3.7&12.0&14.0\\
$T_{e0}$ (MK)&1.25&1.10&1.15\\
$r_{\rm{iso,e}}$ ($R_S$)&8.0&9.0&8.5\\
\hline
$r_c$ ($R_S$)&3.6&7.3&6.2\\
$V_c$ (km s$^{-1}$)&164&115&122\\
$r_A$ ($R_S$)&12.5&12.6&22.9\\
$V_A$ (km s$^{-1}$)&342&190&258\\
\enddata
\tablecomments{Power-law and polytropic indices (top five rows) are obtained by fitting of in situ measurements. 
Initial temperatures and isothermal distances (middle four rows) are set manually to best fit the data. 
Critical points and velocities (bottom four rows) are given by the model.}
\end{deluxetable}

\clearpage


\end{document}